MSCF Investments
# Empirical Project
# Economic Forces in Stock Returns by Yue Chen and Mohan Li

## I. Abstract

When analyzing the components influencing the stock prices, it is commonly believed that economic activities play an important role. More specifically, asset prices are more sensitive to the "systematic" economic news that impose a pervasive effect on the whole market. Moreover, the investors will not be rewarded for bearing "idiosyncratic" risks as such risks are diversifiable. In the paper "Economic Forces and the Stock Market(1986)", the authors introduced an attribution model to identify the specific systematic economic forces influencing the market. They first defined and examined five classic factors from previous research papers: Industrial Production(MP), Unanticipated Inflation(UI), Change in Expected Inflation(DEI), Risk Premia(UPR), and The Term Structure(UTS). By adding in new factors— the Market Indices, Consumptions, and Oil Prices— one by one, they examined each factor's significant contribution to the stock return. The paper concluded that the stock returns are exposed to the "systematic" economic news, and they are priced with respect to their risk exposure. Also, the significant factors can be identified by simply adopting their model. Driven by such motivation, we conduct an attribution analysis based on the general framework of their model to further prove the importance of the economic factors and identify the specific identity of significant factors.

## II. Model Description
Based on the general framework of the model introduced in the paper, we aim at exploring the attributions factors of the stock returns. Unlike the paper choosing an individual stock return as the response variable, We choose the S&P 500 Stock Index Return as the dependent variable since it would better represent the assets facing the "systematic" risks. In particular, we choose SPY as it is traded, more liquid, and has more data on Yahoo Finance. Thus, we construct an attribution model to analyze the ex-post attributions of the S&P 500 Stock Index Return as follows:

$$SPY_t = a_0 + a_{TB} * TB_t + a_{OG} * OG_t + a_{MP} * MP_t + a_{EI} * EI_t$$
$$+ a_{UI} * UI_t + a_{RHO} * RHO_t + a_{UPR} * UPR_t + a_{UTS} * UTS_t + \epsilon_t$$

$TB_t$ is the treasury Bill rate at time t. $OG_t$ is the log relative of the Producer Price Index/Crude Petroleum Series. $MP_t$ is the monthly growth of Industrial Production. $EI_t$ is the expected inflation calculated based on Fama and Gibbons (1984). $UI_t$ is the unexpected inflation at t, calculated by the actual inflation subtracted by expected inflation. $RHO_t$ is the real interest rate (ex-post). $UPR_t$ is the risk premium at t. $UTS_t$ is the term structure at t.

**Data**

Obtaining the data from fred.stlouisfed.org, we decide to use monthly data as it is more suitable to apply OLS under normal assumptions and most of the macroeconomic data are updated monthly. We choose the time period to be the most recent ten years, i.e., 2011-01-01 till now.

**Industrial Production**

To get the MP$_t$, We first obtain the data for the rate of US industrial production, denoted as IP$_t$. Then, we calculate $MP_t = log(IP_t) - log(IP_{t-1})$ to find the monthly growth of IP. As the stock prices involve the valuation of long-term cash flows, simply including MP$_t$ might not capture all the changes in the stock index since the changes in the stock index could reflect multiple monthly changes. As a result, we considered adding in the annual growth of IP, YP$_t$, to account for the long-term effect. However, after examination, this term is highly insignificant, so we exclude it in our final model.

**Inflation**

According to the paper, we define the unexpected inflation $UI_t = I_t - E[I_t|t-1]$ where $I_t$ is the realized monthly first difference in log(Consumer Price Index$_t$) and $EI_t = E[I_t|t-1]$ is the expected inflation modeled based on Fama and Gibbons (1984). We believe that both EI$_t$ and UI$_t$ are responsible for the stock return as they capture different aspects of inflation information. For TB$_t$, we get the monthly data of the Treasury Bill yield. Moreover, we define $RHO_t = TB_{t-1} - I_t$ to be the ex-post real return of the Treasury Bills. These are the factors contributing to the stock return relating to inflation.

To estimate the expected inflation, we adopt a similar approach to that proposed in Fama and Gibbons (1984). The basic idea is to use a time series model on the inflation data to "filter out" trends and leave out the residuals which serve as an indication of unexpected components in inflation. To determine which model to use, we draw the ACF plot of the inflation data (Appendix E). One technique we use here is that of differencing: we see that the inflation series itself is highly auto-correlated with a lag of more than 20, but by differencing the series, we have a much less auto-correlated series, and we draw conclusions from plots of this series. From the plot, we conclude that at least lag 2 of autoregression is significant, thus, we decide to use AR(2) as the model for the data. After fitting the model, we collect the residuals and treat them as the unexpected inflation rate.

**Risk Premia**

We define the risk premia $UPR_t = low - grade\ bond\ return_t - LGB_t$, where the low-grade bond return is the "Baa-rated and under" Bond portfolio return at t and LGB$_t$ is the portfolio return on the long-term government bonds at t. By adopting the UPR$_t$, we hope to capture the effect in the stock return caused by unexpected changes in the risk premia as UPR$_t$ could reflect the changes in the level of risk-aversion and risk premium.

**Term Structure**

We define the term structure at t as $UTS_t = LGB_t - TB_{t-1}$ to capture the changes in the term structure contributing to the changes in stock price.

**Oil Prices**

Commodity prices have always been influential "systematic" factors contributing to stock returns. In order to test this intuition, we choose the oil price as the representative commodity factor. We define $OG_t$ to be the log of Producer Price Index/Crude Petroleum Series.

**Descriptive Features of the Data**

By plotting the scatter plots of Y and X's ([Appendix A](#)) and calculating statistics ([Appendix C](#)) of the data, we find that the features are centered around 0, but they have very different scales in terms of variance, thus, normalization is needed. We also see that there are no obvious outliers in the data. From the plots of features against time ([Appendix B](#)), we see these features are highly non-stationary, and some of them observe clear trends, this motivates one of our future improvements of the project: to use time series models to process the features so that they are more stationary across time.

## III.  Regression Results

We perform the OLS regression using the selected response and independent factors. By normalizing the response and independent variables, we do not need to include the constant term $a_0$. The regression result is shown in [Appendix F](#). The $R^2$ of the model is 0.196, indicating that 19.6% of the variation in data can be explained by the model.

## IV.  Interpretation

The estimated coefficient of **$TB_t$** is -0.8239, showing that an increase in the Treasury Bill yield would decrease the SPY. A higher Treasury Bill Yield would lead more people to invest in bonds and decrease stock prices, equivalent to lower returns. The estimated coefficient of **$OG_t$** is 0.3794, indicating an increase in the log of oil price would cause an increase in the SPY. The energy firms' stock price, especially oil companies, would increase with the increasing oil price. As SPY is composed of the 500 largest publicly-traded companies in the US, energy firms tend to be on the list, thus driving the return of SPY up. The estimated coefficient of **$MP_t$** is -0.2808, signifying an increase in the monthly growth of industrial production would result in a decrease in the SPY return. This result contradicts our normal financial intuition, as normally increasing industrial production would lead to increasing stock prices. We suspect that this counter-intuition result may be due to the effect of excluding the yearly growth factor (**$YP_t$**) as it captures the long-term effect that the industrial production imposing on the stock returns. Further analysis is needed to explore these relationships. The t-statistics for $TB_t$, $OG_t$, $MP_t$ are -2.192, 3.033, -2.470, respectively, which are all statistically significant at a confidence level of 5%. So, we have sufficient evidence to reject the null hypothesis of $a_{TB} = 0$, $a_{OG} = 0$, $a_{MP} = 0$.

For **expected inflation factor $EI_t$** and **unexpected inflation $UI_t$**, the estimated coefficients are 0.2329 and 0.5082, respectively. It is reasonable to believe that the higher the inflation, the higher the stock index return. However, the t-statistics for both factors are smaller than the absolute value of 2. We do not have sufficient evidence to reject the null hypothesis that the coefficients for those two factors are 0. For factor **$RHO_t$**, the estimated coefficient is 0.8130, indicating an increase in the ex-post real return of the Treasury Bills would cause an increase in the SPY return. This real interest rate completely takes out the inflation effect and only accounts for the economic growth. So, when the economic growth improves, the stock market can rise while the real interest rate increases. As the US economy has been booming over the most recent ten years, the coefficient with a positive sign makes sense. However, we cannot reject the null hypothesis because the t-stats is 1.633.

For the factors **UPR**$_t$ and **UTS**$_t$, they both have negative estimated coefficients, which are -0.1837 and -0.0424. These perfectly align with the intuition as an increase in the risk premia or the term structure would cause a decrease in the stock return, because investors tend to withdraw from the stock market when the risk premia and term structure increase. However, the t-stats for both factors are -1.619 and -0.460. We fail to reject the null hypothesis that coefficients are 0.

**Model Validation**

In order to support our argument, we need to perform tests to examine whether the OLS assumptions are valid.

A. Test for Linearity

    The linear model seems appropriate by plotting the actual Y with the fitted values, ([Appendix G](Appendix G)). According to the X's correlation heatmap([Appendix D](Appendix D)), there is no strong correlation among X's. Besides, the residuals vs. fitted values plot ([Appendix I](Appendix I)) show random patterns, indicating such model would be a good fit.

B. Normality of the error terms

    Based on the Q-Q plot with distance set to Normal distribution ([Appendix H](Appendix H)), we can conclude that the residuals are approximately normal.

C. No multicollinearity among the predictors

    During the selection of features, we make sure no perfectly correlated ones are selected together, this is backed by the fact of no multicollinearity warning during regression. The condition is satisfied.

D. No autocorrelation of the Error terms

    By performing the Durbin-Watson Test ([Appendix J](Appendix J)), we found the test statistics is 2.42865, indicating little to no autocorrelation. Thus, the assumption is satisfied.

E. Test for homoscedasticity

    There is no obvious sign of heteroskedasticity from the plot of the residuals and fitted values ([Appendix I](Appendix I)), since the thickness of the cluster is roughly the same along the x-axis.

**Possible Improvements**

Some possible improvements for our model: (1). We could include more macroeconomic factors, such as GDP and other types of consumption index. (2). A simple linear model might not be enough for time-series data. We could consider other more advanced models such as ARIMA to incorporate lagged terms.

# V.   Conclusion

This project follows the framework proposed by Chen, et al. (1986) to examine the systematic influences of macro variables on stock returns. By efficient market theory, asset prices should depend on their exposure to the state variables which describe the economy.

We first choose a set of variables based on the paper and their economic meanings. In the regression, some of them are found to be significant, we give interpretations for these variables. To determine unexpected components in certain variables, we adopt a time series model. We test the model on the most recent data with SPY ETF. Our conclusion is that stock returns are exposed to systematic economic news, and there are certain variables that are significant under the linear model framework.

## VI.   Reference

- Chen, N.-F., Roll, R., & Ross, S. A. (1986). Economic Forces and the Stock Market. The Journal of Business, 59(3), 383–403. http://www.jstor.org/stable/2352710
- Fama, Eugene F. and Gibbons, Michael R., (1984), A comparison of inflation forecasts, Journal of Monetary Economics, 13, issue 3, p. 327-348, https://EconPapers.repec.org/RePEc:eee:moneco:v:13:y:1984:i:3:p:327-348

## VII.   Appendix

- <u>Appendix A - Scatterplots of Y vs. X's</u>

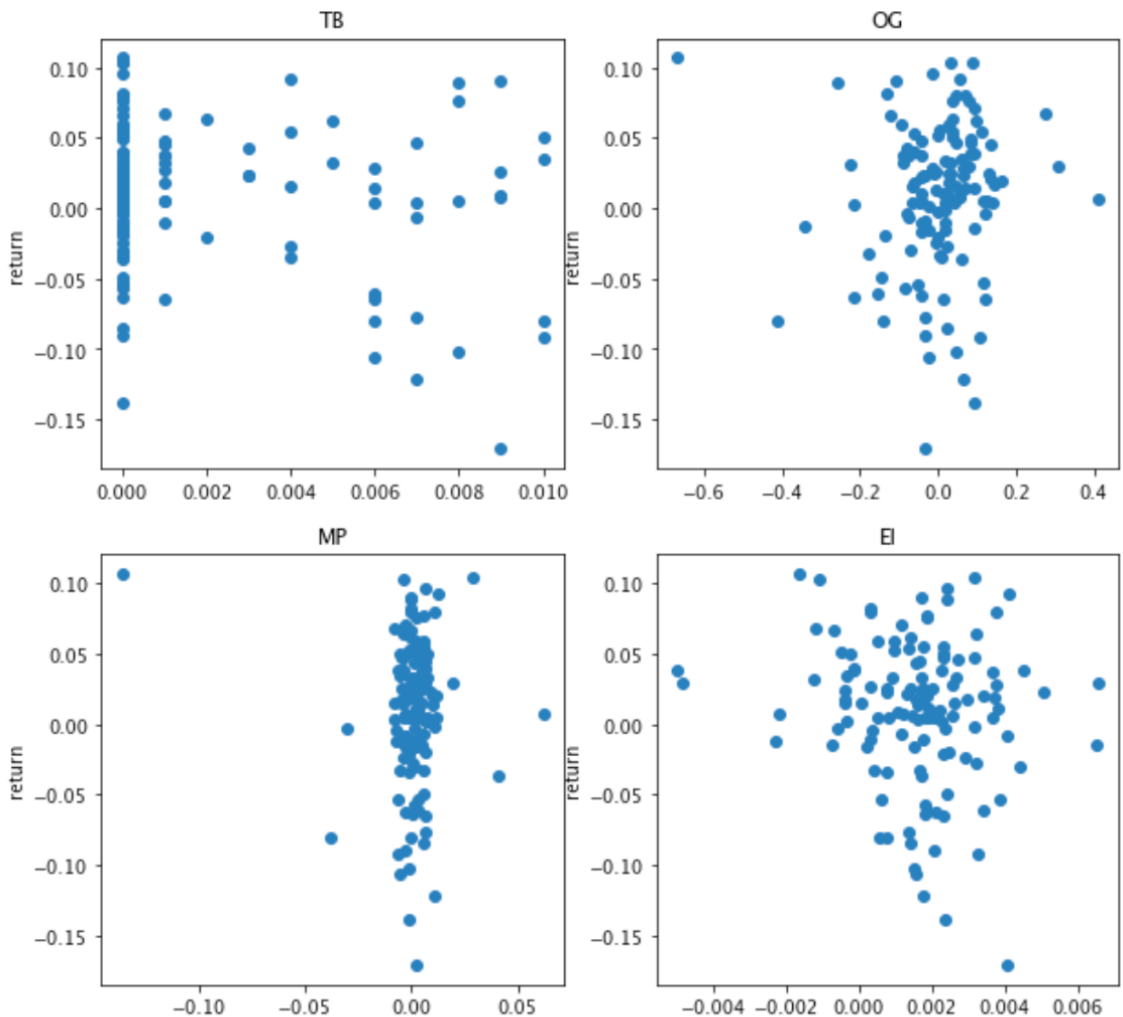

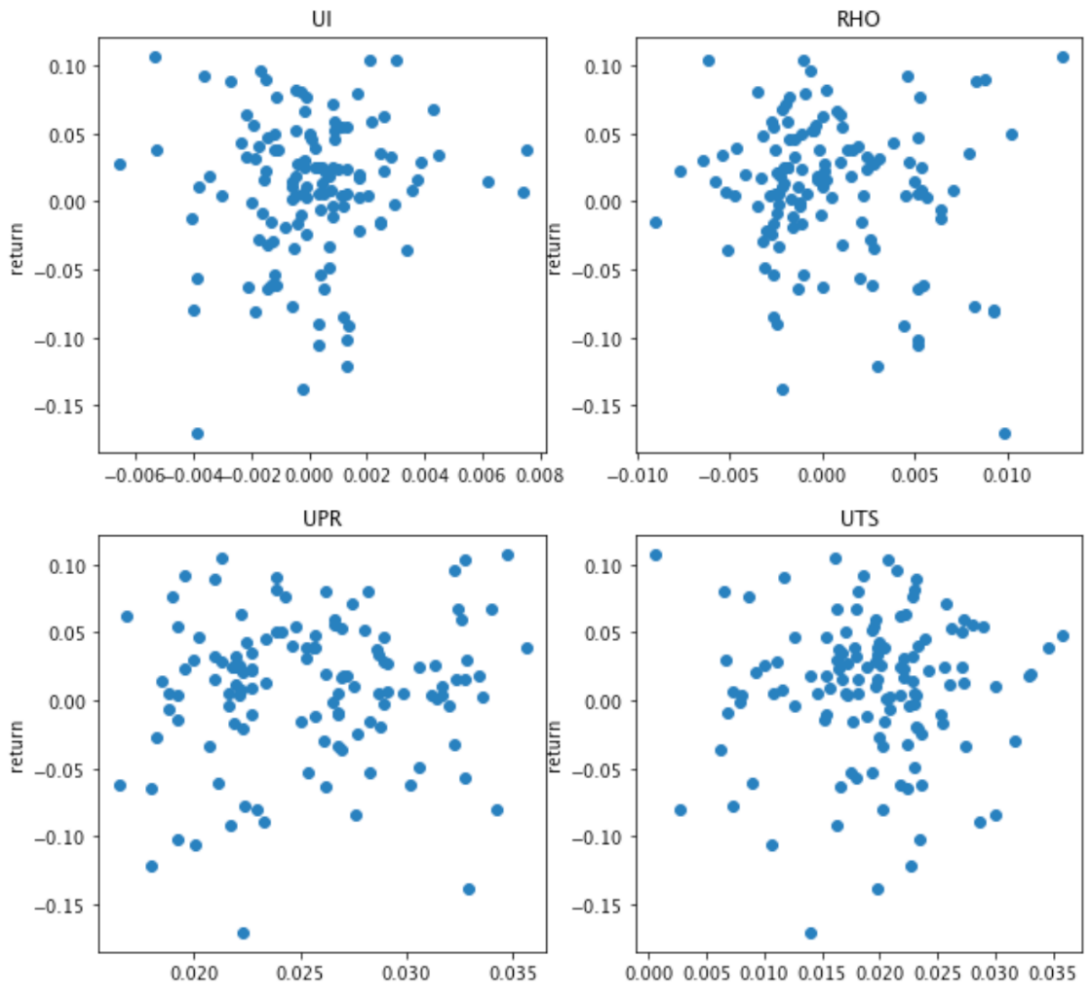

- Appendix B - Time Plots of Y and X's
  Y:

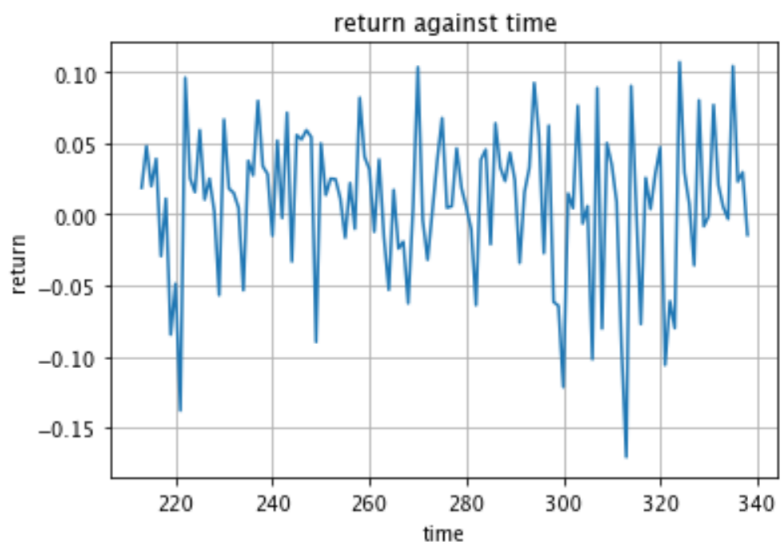

X:

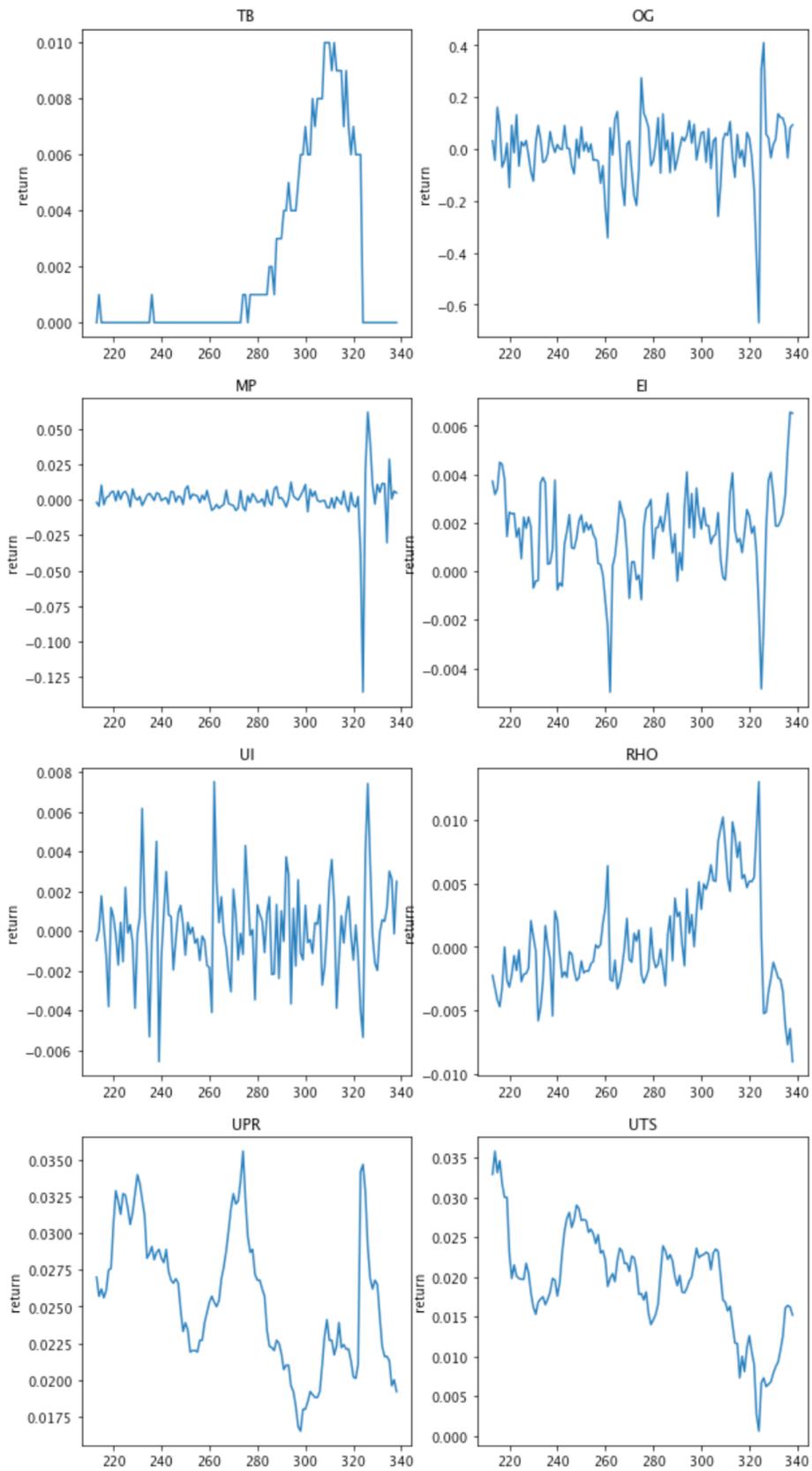

- Appendix C - Basic Descriptive Features of the Data

|  | TB | OG | MP | EI | UI | RHO | UPR | UTS |
|---|---|---|---|---|---|---|---|---|
| count | 126.000000 | 126.000000 | 126.000000 | 126.000000 | 126.000000 | 126.000000 | 126.000000 | 126.000000 |
| mean | 0.002032 | -0.001740 | 0.000676 | 0.001581 | 0.000056 | 0.000403 | 0.025506 | 0.019354 |
| std | 0.003165 | 0.124377 | 0.015629 | 0.001764 | 0.002279 | 0.004070 | 0.004638 | 0.006561 |
| min | 0.000000 | -0.668904 | -0.135928 | -0.004981 | -0.006576 | -0.009008 | 0.016500 | 0.000600 |
| 25% | 0.000000 | -0.045208 | -0.003154 | 0.000654 | -0.001303 | -0.002374 | 0.022000 | 0.016225 |
| 50% | 0.000000 | 0.015265 | 0.001030 | 0.001730 | -0.000025 | -0.000605 | 0.025500 | 0.019800 |
| 75% | 0.004000 | 0.063740 | 0.005289 | 0.002433 | 0.001178 | 0.002728 | 0.028775 | 0.023000 |
| max | 0.010000 | 0.410835 | 0.061959 | 0.006564 | 0.007512 | 0.012990 | 0.035600 | 0.035800 |

- Appendix D - X's Correlations Heatmap

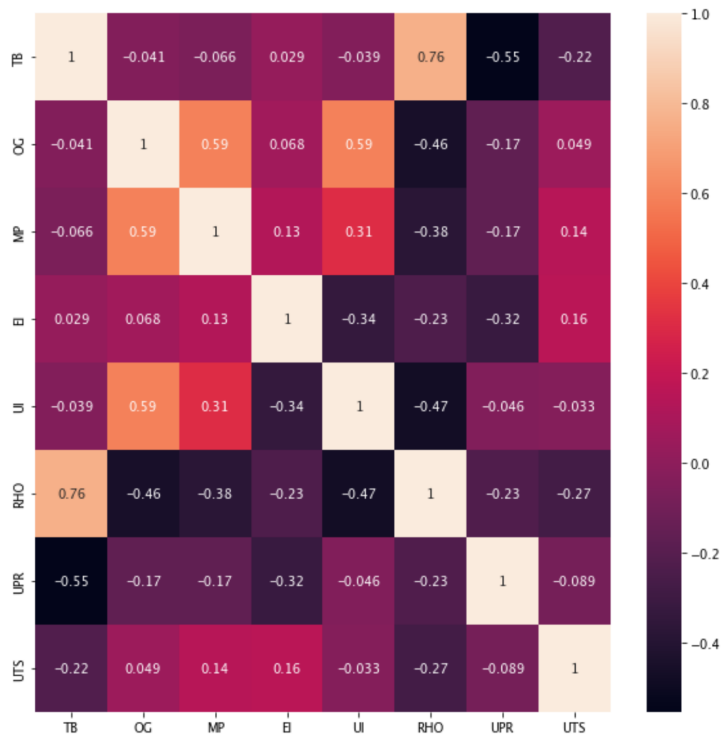

- Appendix E - ACF Plots

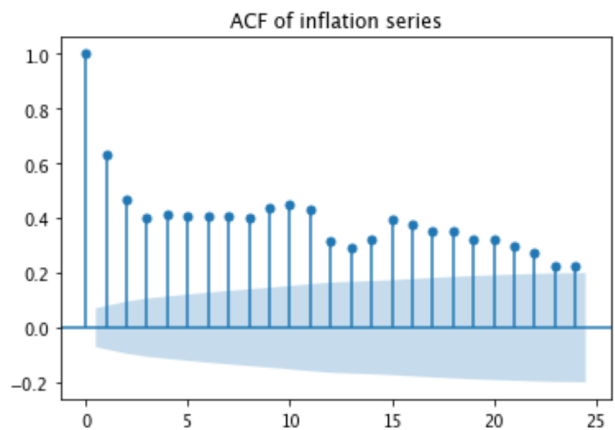

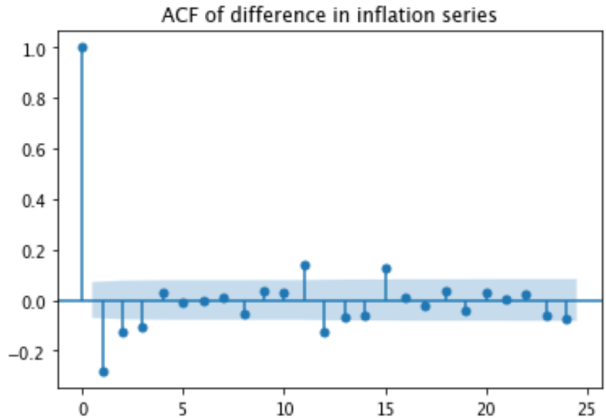

- Appendix F - OLS Regression Result

```
Regression results:
                            OLS Regression Results
==============================================================================
Dep. Variable:                    ret   R-squared (uncentered):                   0.196
Model:                            OLS   Adj. R-squared (uncentered):              0.142
Method:                 Least Squares   F-statistic:                              3.601
Date:                Mon, 11 Oct 2021   Prob (F-statistic):                    0.000893
Time:                        01:51:26   Log-Likelihood:                         -164.52
No. Observations:                 126   AIC:                                      345.0
Df Residuals:                     118   BIC:                                      367.7
Df Model:                           8
Covariance Type:            nonrobust
==============================================================================
                 coef    std err          t      P>|t|      [0.025      0.975]
------------------------------------------------------------------------------
TB            -0.8239      0.376     -2.192      0.030     -1.568      -0.080
OG             0.3794      0.125      3.033      0.003      0.132       0.627
MP            -0.2808      0.114     -2.470      0.015     -0.506      -0.056
EI             0.2329      0.235      0.993      0.323     -0.231       0.697
UI             0.5082      0.299      1.701      0.092     -0.083       1.100
RHO            0.8130      0.498      1.633      0.105     -0.173       1.799
UPR           -0.1837      0.113     -1.619      0.108     -0.408       0.041
UTS           -0.0424      0.092     -0.460      0.646     -0.225       0.140
==============================================================================
Omnibus:                        3.182   Durbin-Watson:                     2.429
Prob(Omnibus):                  0.204   Jarque-Bera (JB):                  3.455
Skew:                           0.009   Prob(JB):                          0.178
Kurtosis:                       3.811   Cond. No.                           14.0
==============================================================================

Warnings:
[1] Standard Errors assume that the covariance matrix of the errors is correctly specified.
```

- Appendix G - Actual Y vs. Fitted Values of Y

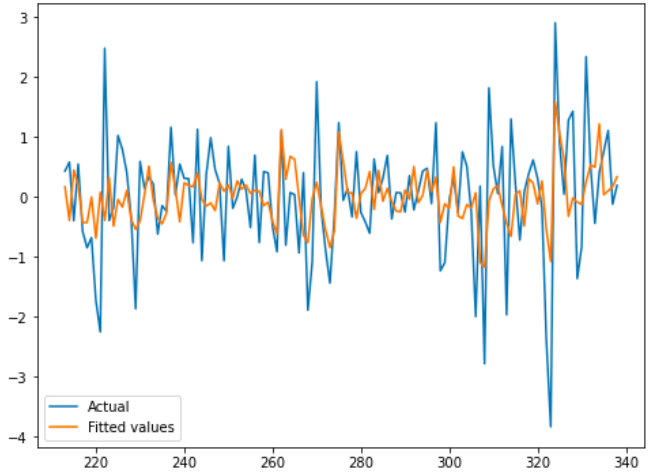

- Appendix H - QQ Plot of the Residuals

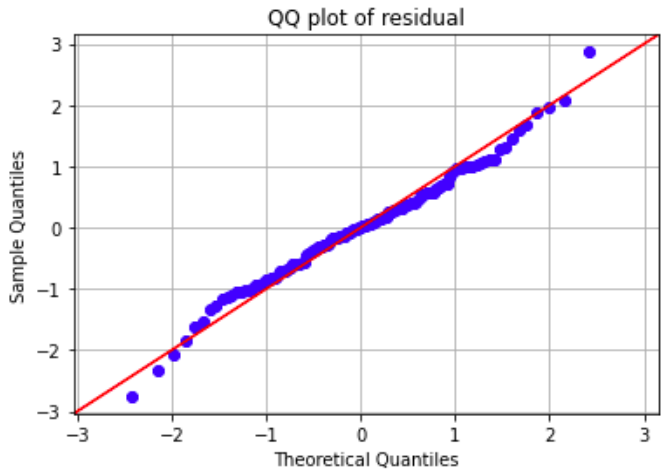

- Appendix I - Residuals vs. Fitted Values of Y

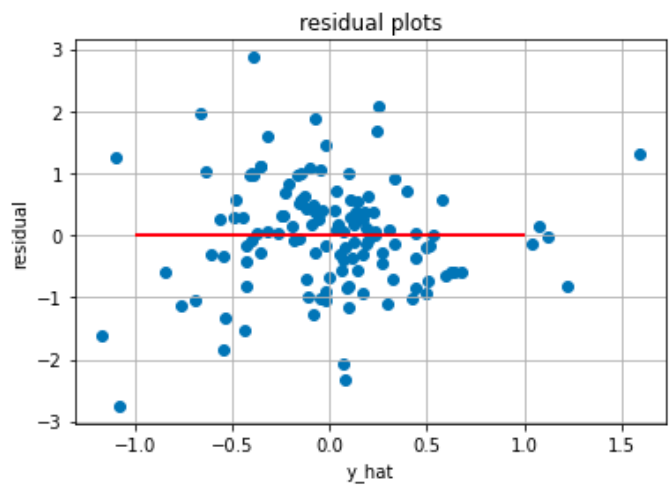

- <u>Appendix J - Durbin-Waston Test Result</u>

```
Performing Durbin-Watson Test
Values of 1.5 < d < 2.5 generally show that there is no autocorrelation in the data
0 to 2< is positive autocorrelation
>2 to 4 is negative autocorrelation
-------------------------------------
Durbin-Watson: 2.4286512923073307
Little to no autocorrelation

Assumption satisfied
```